\documentclass[aps,pre,twocolumn,showpacs,amssymb,superscriptaddress]{revtex4-1}
\usepackage{graphicx}
\usepackage{amsfonts,amsmath}

\newcommand{\argmin}{\operatorname{argmin}}

\begin{document}


\title{Inferring the spatiotemporal DNA replication program from noisy data}



\author{A. Baker}
\author{J. Bechhoefer}
\affiliation{Department of Physics, Simon Fraser University, Burnaby, British Columbia, V5A 1S6, Canada.}

\date{\today}

\begin{abstract}

We generalize a stochastic model of DNA replication to the case where replication-origin-initiation rates vary locally along the genome and with time.  Using this generalized model, we address the inverse problem of inferring initiation rates from experimental data concerning replication in cell populations.  Previous work based on curve fitting depended on arbitrarily chosen functional forms for the initiation rate, with free parameters that were constrained by the data.  We introduce a non-parametric method of inference that is based on Gaussian process regression.  The method replaces specific assumptions about the functional form of initiation rate with more general prior expectations about the smoothness of variation of this rate, along the genome and in time.  Using this inference method, we recover,  with high precision, simulated replication schemes from noisy data that are typical of current experiments.

\end{abstract}



\maketitle


\section{Introduction}
\label{sec:intro}

Cells must accurately duplicate their DNA content at every cell cycle.  Depending on the organism, DNA replication can initiate at one or at multiple sites called \textit{origins of replication}.  The DNA is copied by a pair of oppositely moving \textit{replication forks} that propagate away from each origin.  These forks actively copy the genome away from the origin until they encounter another replication fork.  DNA replication can thus be modeled as a process of initiation, growth, and coalescences occurring in an asynchronous, parallel way until the whole genome is copied.  In this process, initiation has been observed to be a stochastic process \cite{aFriedmanGenesCells1997,herrick00,lucas00,aPatelMolBiolCell2006,aRhindNatCellBiol2006,aCzajkowskyJMolBiol2008}, while fork propagation, at the large scales (10--100 kb) between origins, is largely deterministic, and often  constant \cite{sekedat10}.

The elements of stochastic initiation, deterministic growth, and coalescence are formally equivalent to the processes of nucleation, growth, and coalescence in crystallization kinetics, and this equivalence has inspired efforts to model DNA replication kinetics using the formalism developed in the 1930s by Kolmogorov, Johnson, Mehl, and  Avrami (KJMA) for crystallization kinetics \cite{aKolmogorovBullAcadSciURSS1937,*aJohnsonTransAIME1939,*aAvramiJChemPhys1939,*aAvramiJChemPhys1940,*aAvramiJChemPhys1941}.  Of course, DNA replication takes place in a space that is topologically one dimensional, a fact that allows one to take advantage of exact solutions to the KJMA equations in one dimension \cite{sekimoto91,*bennaim96}.

The rate of initiation of origins is typically highly variable, both in space, along the genome, and in time, throughout \textit{S phase}, the part of the cell cycle in which the genome is duplicated.  In many cases, we can describe the initiation process by a rate $I(x,t)$, where $I(x,t) \, dx \, dt$ gives the probability of initiation to occur in $(x,x+dx)$ at $(t,t+dt)$ given that $x$ is unreplicated up until time $t$.  Loosely, we will say that $I(x,t)$ is the probability for an origin to initiate, or ``fire" at $(x,t)$.  

In addition to its intrinsic theoretical interest, describing replication stochastically can help biologists understand better the biological dynamics underlying replication.  As we discuss below, experiments have recently begun to deliver large amounts of data concerning cell populations undergoing replication.  For example, it is now possible to measure the fraction of cells $f(x,t)$ that have replicated the locus $x$ along the genome by a time $t$ after the beginning of S phase \cite{aRaghuramanScience2001}.  In contrast to the case of crystallization kinetics, there is little fundamental understanding of the structure of the initiation function $I(x,t)$.  Since direct observation of initiations \textit{in vivo} has not been possible, the task is to estimate, or infer, $I(x,t)$ from data such as the replication fraction $f(x,t)$ or---more conveniently, it will turn out---the unreplicated fraction $s(x,t) = 1-f(x,t)$, which is also the probability that the locus $x$ is unreplicated at time $t$.

In this paper, we have two goals:  the first, presented in Sections~\ref{sec:gen-rep} and~\ref{sec:indep-ori-firing}, is to collect and generalize previous results on the application of the KJMA formalism to DNA replication.  Previous work has focused on special cases:  models of replication in \textit{Xenopus laevis} (frog) embryos were based on experiments that averaged data from the whole genome \cite{herrick02} and thus could neglect spatial variations.  Conversely, in recent experiments on a small section of a mouse genome, spatial variations dominated and temporal variations could be neglected.  In budding yeast, origins are restricted to specific sites along the genome  \cite{aNieduszynskiNucleicAcidsRes2007}, which also leads to a restricted form of the initiation function.  In general, however, both spatial and temporal variations are important, and we extend here the full KJMA formalism to handle such cases.  Section~\ref{sec:example-rep-prog} gives a brief example that illustrates the kinds of results and insights that this approach to modeling replication can provide.

The second goal is to present a new way to infer initiation rates $I(x,t)$ from replication data such as $s(x,t)$.  Replication timing data are increasingly available for a variety of organisms and cell types \cite{aRaghuramanScience2001,schubeler02,aHirataniPLoSBiol2008,aHansenProcNatlAcadSciUSA2010,muller13}, and advances in experimental techniques now allow the determination of the probability distribution of genome-wide  replication timing at fine spatial and temporal scales.  For instance, in yeast, the unreplicated fraction profiles have been determined at 1 kb resolution in space and 5 min resolution in time \cite{hawkins13}.  The increasing availability of data makes the ability to infer initiation rates important.  

Our main result, presented in Section~\ref{sec:inferring-local-init-rate}, is to adapt the technique of \textit{Gaussian process regression} to ``invert" experimental replication data and estimate the initiation function $I(x,t)$ and  fork velocity $v$.  Previous approaches have mainly used curve fitting, a technique that postulates a suitable functional form for $I(x,t)$, with free parameters that are then constrained by fitting to the data.  This technique was used to infer initiation functions in frog embryos \cite{herrick02}, budding yeast  \cite{aMouraNucleicAcidsRes2010,aLuoBMCBioinformatics2010,aYangMolSystBiol2010,hawkins13}, and limited regions of human somatic cells \cite{demczuk12}.

Although the above efforts were successful, curve-fitting methods are time consuming, requiring considerable effort to generate initial guesses that are close enough to the final inference.  The situation is even more difficult if one wants to describe replication over the whole genome of higher eukaryotes.  In these organisms, initiations are not limited to well-positioned replication origins but also occur in large extended initiation zones whose functional form is not known \textit{a priori}.  Furthermore, the mapping of well-positioned replication origins and extended initiation zones along the genome is difficult \cite{aGilbertNatRevGenet2010}, and not much is known about the firing-time distributions.  These added uncertainties make curve-fitting approaches to local genomic data in higher eukaryotes problematic.  

Given the difficulty of extending and automating curve-fit approaches, we explore here an alternative that does not depend on knowing \textit{a priori} the functional form of the initiation function.  The technique, Gaussian process regression, is based on the Bayesian approach to data analysis and gives a systematic way to infer the initiation rate without making detailed assumptions about its functional form in the way required of curve-fit methods.  Although Gaussian process regression is more powerful than curve-fitting methods, it can be simpler to apply.  Because no detailed tuning of initiation conditions is required, the method can in principle be automated.  In contrast, curve-fitting methods require a good technical understanding to use successfully.

\section{General replication program}
\label{sec:gen-rep}

We begin by establishing relationships that must be obeyed by any spatiotemporal replication program with a constant fork velocity.  We then show that many quantities of interest, such as the densities of right- and left-moving forks or the initiation and termination densities, are related to derivatives of the unreplicated fraction profiles.  Then we describe briefly how to use these relationships to characterize the replication program.  

\subsection{DNA replication kinetics quantities}
\label{sec:definitions}

\begin{figure}[hbt]
	\includegraphics[width=0.8\linewidth]{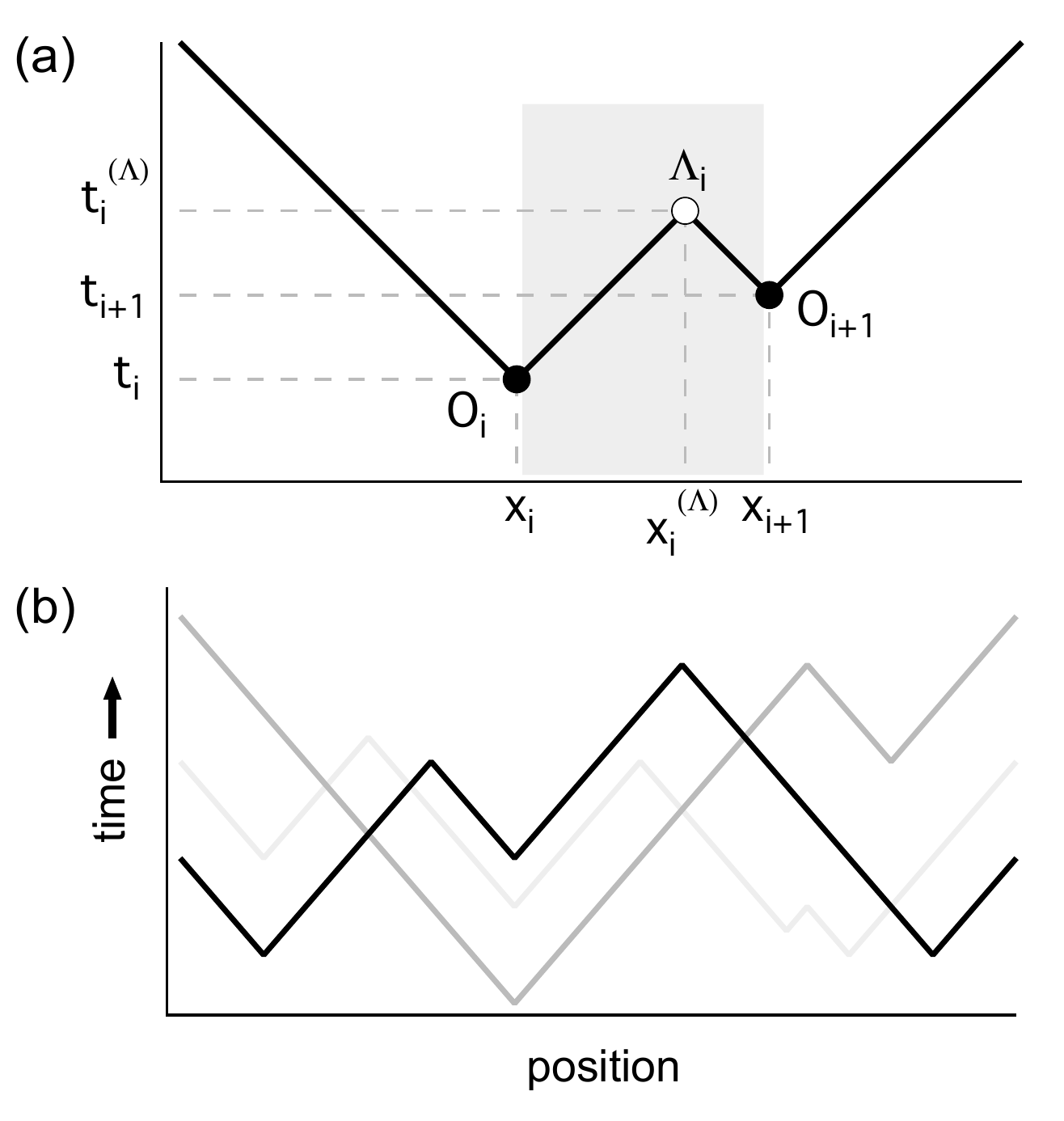}
	\caption{\label{fig:replication_program}
Spatiotemporal representation of the replication program.  (a) Replication program in one cell cycle. From each replication origin $O_i$ (filled disk), two replication forks propagate at constant velocity $v$ until they meet a fork of the opposite direction at a replication terminus $\Lambda_i$ (hollow disk). The \textit{replication timing curve}---the time at which a locus is replicated---is given by the intersecting set of propagation lines of the replication forks (dark zig-zag line).  The shaded area shows the domain of terminus $\Lambda_i$.   (b) Replication program in several cell cycles.  The number of activated origins, their genomic positions, and firing times change from one cell cycle to another.}
\end{figure}

If the replication fork velocity $v$ is constant, the replication program in one cell cycle is completely specified by the genomic positions and firing times of the replication origins.  From each origin, two divergent forks propagate at constant velocity until they meet and coalesce with a fork of the opposite direction at a replication terminus [Fig.~\ref{fig:replication_program}(a)]. The spatial and temporal coordinates of replication termini, as well as the propagation lines of the replication forks, and the replication timing (the time at which a locus is replicated) can all be derived from the genomic positions and firing times of the replication origins. Note that the inherent stochasticity of the replication program implies that the number of activated origins, along with their positions and firing times, change from one cell cycle to another, as depicted in Fig.~\ref{fig:replication_program}(b). Consequently, the number of terminations and initiations, the number of forks, and the replication timing curve all change from one cell cycle to another.

Let us define several quantities describing a stochastic DNA replication program.  The \textit{initiation and termination densities}, $\rho_{\rm init}(x,t)$ and $\rho_{\rm ter}(x,t)$, give the (ensemble average) number of initiation and termination events observed in any given spatiotemporal region. The corresponding spatial densities are given by $\rho_{\rm init}(x) = \int_{0}^{t_\infty} dt \, \rho_{\rm init}(x,t)$ and $\rho_{\rm ter}(x) = \int_{0}^{t_\infty} dt \, \rho_{\rm ter}(x,t)$.  Note that although the integration formally is to $t=\infty$, the end of replication for a finite genome of length $L$ will at a finite (but stochastic) time $t_{\rm end}$ \cite{bechhoefer07,*yang08}.  Often, $\rho_{\rm init}(x)$ is called the \textit{efficiency} of the locus $x$, as it equals the fraction of cells where locus $x$ has initiated.

In this paper, we use the compact notation $(\pm)$ to distinguish right-moving forks (velocity $+v$) from left-moving forks (velocity $-v$). The \textit{fork densities} $\rho_{\pm}(x,t)$ give the spatial densities of $(\pm)$ forks at a given time $t$. In other words, the (ensemble average) number of $(\pm)$ forks in a genomic region $[x_1,x_2]$ at time $t$ is given by $\int_{x_1}^{x_2} dx \, \rho_{\pm}(x,t)$. Also, as forks propagate at velocity $\pm v$, the number of $(\pm)$ forks crossing the locus $x$ during $[t_1,t_2]$ is given by $\int_{t_1}^{t_2} v dt \, \rho_{\pm}(x,t)$. Consequently, the proportions of cell cycles where the locus $x$ is replicated by a $(\pm)$ fork is given by $p_{\pm}(x) = \int_{0}^{t_\infty} v dt \, \rho_{\pm}(x,t)$. The \textit{replication fork polarity} $p(x) = p_{+}(x)-p_{-}(x)$ measures the average directionality of the fork replicating the locus $x$.

\textit{Replication timing}---the time when a locus is replicated---changes from one cell cycle to another.  The variations can be intrinsic, due to stochastic initiation in an individual cell, and extrinsic, due to a population of cells.  These variations lead to a probability distribution $P(x,t)$ for the replication timing at locus $x$.  The closely related \textit{unreplicated fraction} $s(x,t)$ is defined to be the fraction of cells where $x$ is unreplicated at time $t$.  Since $s(x,t)$ equals the probability that replication at $x$ occurs after $t$, we see that $P(x,t) = -\partial_t s(x,t)$. The ensemble average of the replication timing, or \textit{mean replication timing}, is then \footnote{
Our definition differs from that of \cite{aRetkutePhysRevLett2011,aRetkutePhysRevE2012} in that we neglect the very small probability that no initiations occur on a chromosome.  Replication is then not well defined.}
\begin{align}
	T(x) = \int_{0}^{t_\infty} dt \, P(x,t) \, t = \int_{0}^{t_\infty} dt \, s(x,t) \,.
\end{align}

\subsection{Derivatives of the unreplicated fraction profiles}
\label{sec:derivatives}

We can establish a number of relations among the quantities defined in Section~\ref{sec:definitions}.  In particular,
\begin{subequations}
\begin{align}
\label{eq:ds_dt}
	v[\rho_+(x,t) + \rho_-(x,t)] &= - \partial_t s(x,t) \,,\\[3pt]
\label{eq:ds_dx}
	\rho_+(x,t) - \rho_-(x,t) &= \partial_x s(x,t) \,,\\[3pt]
\label{eq:rho_pm}
	\rho_\pm(x,t) &= -\tfrac{1}{2} \left( \tfrac{1}{v} \partial_t \mp \partial_x \right) s(x,t) \,, \\[3pt]  
\label{eq:dalembertian_s}
	\rho_{\rm init}(x,t) - \rho_{\rm ter}(x,t) &= -\tfrac{1}{2} v\square \, s(x,t) \,, \\[3pt]
\label{eq:dT_dx}
	p(x) &= vT'(x) \,, \\[3pt]
\label{eq:d2T_dx2}
	\rho_{\rm init}(x) - \rho_{\rm ter}(x) &= \tfrac{1}{2} vT''(x) = \tfrac{1}{2} p'(x) \,,
\end{align}
\end{subequations}
where $\square = \frac{1}{v^2} \partial_t^2 - \partial_x^2$ is the d'Alembertian operator.  See the Appendix for a proof of these relations. 

From Eq.~\eqref{eq:rho_pm}, the densities of right- and left-moving forks are directly given by derivatives of the unreplicated fraction.  The sum of the fork densities in Eq.~\eqref{eq:ds_dt} is related to $P(x,t) = - \partial_t s(x,t)$, the probability distribution of  replication timing at locus $x$. Equations~\eqref{eq:dT_dx} and \eqref{eq:d2T_dx2}, previously derived in \cite{aBakerPLoSComputBiol2012,aAuditNatureProt2013}, and, in special cases, in \cite{aRetkutePhysRevLett2011,aRetkutePhysRevE2012}, show that the shape of the mean replication timing curve $T(x)$ gives direct information about the fork polarity and the relative densities of initiation and termination in a region.  For instance, the replication fork polarity profile $p(x)$ was estimated in the human genome using Eq.~\eqref{eq:dT_dx} and shown to be the key determinant of the compositional and mutational strand asymmetries generated by the replication process \cite{aBakerPLoSComputBiol2012,aBakerEurPhysJE2012a,aBakerEurPhysJE2012b}.   

Contrary to intuition \cite{aRaghuramanScience2001}, the above equations show that there need not be a direct correspondence between well-positioned replication origins and timing-curve minima \cite{aRetkutePhysRevLett2011}. Around a fixed, isolated origin $i$ located at position $x_i$, the initiation density profile reduces to a Dirac delta function: $\rho_{\rm init} = E_i \, \delta(x-x_i)$, where the height $E_i$ is the observed efficiency of origin $i$ (the fraction of cells where origin $i$ has initiated). Equation~\eqref{eq:d2T_dx2} shows that the isolated origin $i$ produces a jump discontinuity of height $2E_i$ in the fork polarity profile. Equation~\eqref{eq:dT_dx} shows that at a minimum in $T(x)$, the fork polarity $p(x)$ must change sign. Mathematically, the efficiency $E_i$ of the origin may or may not be large enough to produce a sign shift in $p(x)$ corresponding to a  minimum of the $T(x)$ curve.  More intuitively, a weak origin (one that rarely fires in a cell cycle) in a region that is almost always replicated by a nearby strong origin may not affect the timing curve enough to produce a local minimum.  As a result, even fixed, isolated origins do not necessarily imply minima in the mean replicating time curve \cite{aRetkutePhysRevLett2011}.  Indeed, in budding yeast, about one origin in three is not associated with a local minimum of the timing curve \cite{aYangMolSystBiol2010}.

\section{Independent origin firing}
\label{sec:indep-ori-firing}

The results of Section~\ref{sec:derivatives} are valid for any initiation rule.  If, also, origins fire independently, then the whole spatiotemporal replication program is analytically solvable.  ``Independence'' here means that an initiation event neither impedes nor favors origin initiation at another loci and implies that we can define a local initiation rate of unreplicated DNA, $I(x,t)$.  The local initiation rate then completely specifies the stochastic replication program.  Most models of the replication program proposed so far \cite{aLygerosProcNatlAcadSciUSA2008,aBlowEMBORep2009,aMouraNucleicAcidsRes2010,aLuoBMCBioinformatics2010,aYangMolSystBiol2010,aRetkutePhysRevLett2011} assume the independent firing of replication origins and are thus special cases of the general formalism presented here.  (An exception is \cite{aJunCellCycle2004}.)    The replication program is then formally analogous to a one-dimensional nucleation-and-growth process with time- and space-dependent nucleation/initiation rate.   In the 1930s, the kinetics of nucleation-and-growth processes were analytically derived for crystallization by Kolmogorov, Johnson, Mehl and Avrami in the \textit{KJMA theory} of phase transition kinetics \cite{aKolmogorovBullAcadSciURSS1937,aJohnsonTransAIME1939,aAvramiJChemPhys1939,aAvramiJChemPhys1940,aAvramiJChemPhys1941}.  Here, we will prove that the quantities describing DNA replication---the unreplicated fraction profiles and the probability distribution of the replication timing curve, the density of initiation and termination and of forks---can all be analytically derived from the local initiation rate.

The KJMA formulation of the replication program is an exactly solvable model, as all higher-moment correlation functions can also be analytically derived, for example the joint probability distribution of replication timing at different loci, or the joint densities of initiations at different loci.  We will show that, even when origins fire independently, the propagation of forks creates correlations in nearby replication times and in nearby initiation events.  

Many of these relationships were previously derived for the special case of well-positioned replication origins \cite{aYangMolSystBiol2010,aRetkutePhysRevE2012}.  The present formalism is more general, as it can include extended initiation zones,  and offers a more compact and elegant derivation of these relationships.

\subsection{Unreplicated fraction}

\begin{figure}[hbt]
	\includegraphics[width=0.8\linewidth]{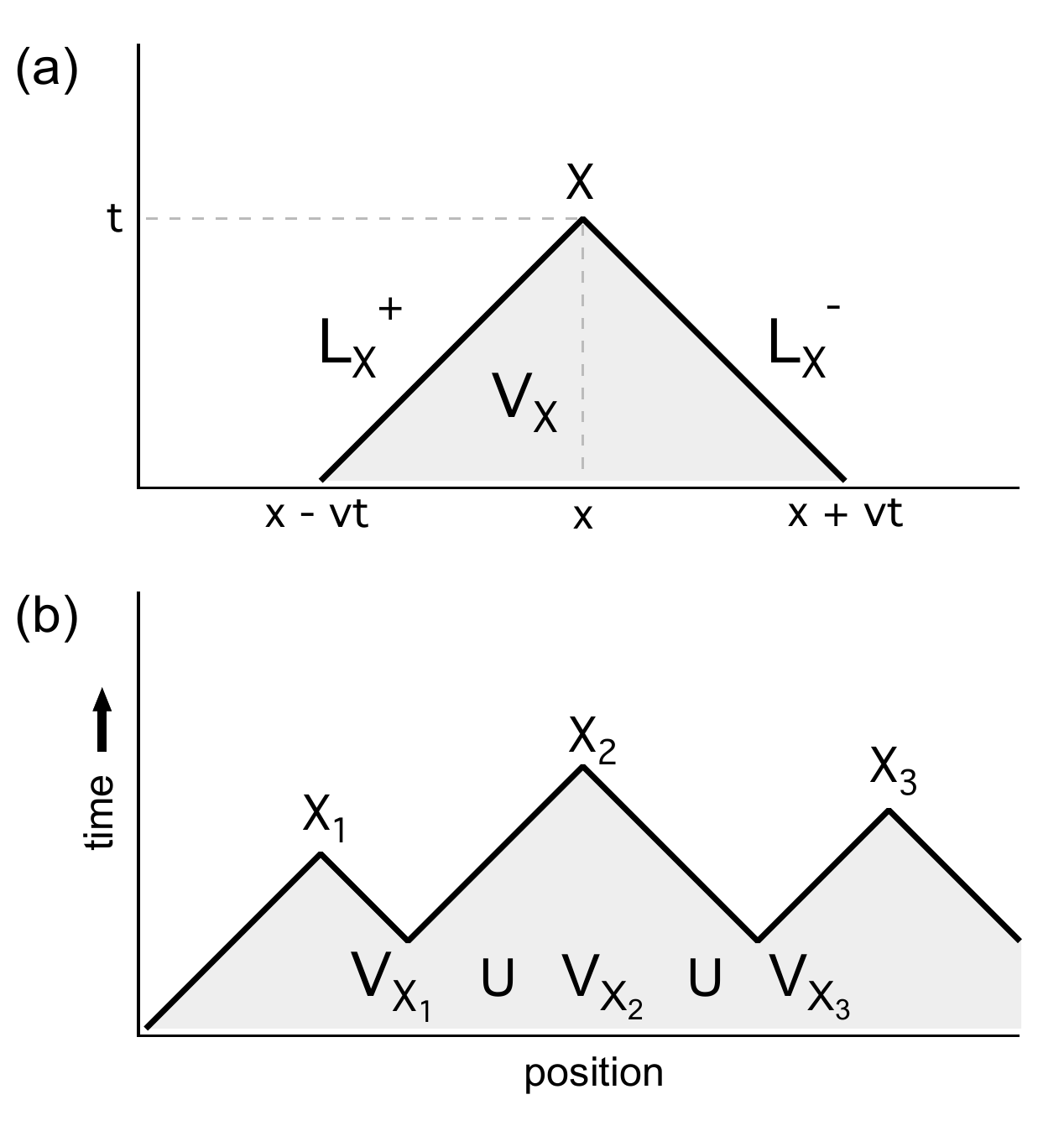}
	\caption{\label{fig:KJMA}
	Kolmogorov's argument. (a) A locus $x$ is unreplicated at time $t$ iff no initiation occurs in the past cone $V_X$ of $X=(x,t)$, the gray region demarcated by the lines $L_X^{\pm}$.  (b) The loci $x_1,x_2,x_3$ are all unreplicated at times $t_1,t_2,t_3$ iff no initiation occurs in $V_{X_1}\cup V_{X_2}\cup V_{X_3}$ (gray region).}
\end{figure}

We first note that the locus $x$ is unreplicated at time $t$ if and only if (iff) no initiations occur in the past ``cone'' $V_{(x,t)}[v]$ of $(x,t)$ [gray area in Fig.~\ref{fig:KJMA}(a)] defined by
\begin{equation}
	V_{(x,t)}[v] = \{ (x',t'):\; | x - x' |\leq v (t-t') \} \,.
\label{eq:past-cone}
\end{equation}
When the context is unambiguous, we will use the more compact notation $X=(x,t)$ and $V_X \equiv V_{(x,t)}[v]$.  The unreplicated fraction then equals the probability that no initiations occur in $V_X$ (Kolmogorov's argument \cite{aKolmogorovBullAcadSciURSS1937}).  As initiations occur independently with an initiation rate $I(x,t)$, this probability is  given by a Poisson distribution with time- and space-dependent rate \cite{VanKampen2007}.  Thus, the unreplicated fraction is given by \cite{aJunPhysRevE2005}
\begin{equation}
	s(x,t) = e^{-\int_{V_X} dt' \, dx' \, I(x',t') } \,.
\label{eq:s_KJMA}
\end{equation}

\subsection{Replication timing and fork densities}

We can extend Kolmogorov's argument to find the fork densities.  From Eqs.~(\ref{eq:rho_pm}) and (\ref{eq:s_KJMA}), we find
\begin{equation}
	\rho_\pm (x,t) = \left[ \int_{L_X^\pm} I \right] s(x,t) \,,
\label{eq:rho_pm_KJMA}
\end{equation}
where the integrals of $I$ over the lines $L_X^+$ and $L_X^-$ in Fig.~\ref{fig:KJMA}(a) are defined as
\begin{align}
        \int_{L_X^\pm} I &\equiv \int_0^t \, dt' \, I[x \mp v(t-t'), t'] \,.   
\end{align}
The interpretation of Eq.~\eqref{eq:rho_pm_KJMA} is straightforward: a $(\pm)$ fork passes by $x$ at time $t$ iff no initiation occurs in $V_X$ and one initiation occurs along $L_X^\pm$.

Similarly, from Eq.~\eqref{eq:ds_dt}, 
\begin{align}
	P(x,t) &=  v\left[ \rho_+ (x,t)+ \rho_- (x,t) \right] 
		= -\partial_t s(x,t)  \nonumber \\[3pt]
	&= v \left[ \int_{L_X^+} I + \int_{L_X^-} I \right] s(x,t) \,.
\label{eq:P_KJMA}
\end{align}
In words:  to have replication at $X=(x,t)$, no initiation occurs in $V_X$ and an initiation  along either the line $L_X^+$ or the line $L_X^-$ causes a fork of velocity $v$ to sweep by.

\subsection{Initiation and termination densities}

The initiation rate $I(x,t)$ gives the number of initiations at an unreplicated site.  The initiation density $\rho_{\rm init} (x,t)$ is then determined by the rate of initiation at $(x,t)$ times the probability that no initiations occurred previously in the triangular area $V_X$ defined in Fig.~\ref{fig:KJMA}(a):
\begin{equation}
	\rho_{\rm init} (x,t) = I(x,t) s(x,t) \,.
\label{eq:rho_ini_KJMA}
\end{equation}

From Eqs.~(\ref{eq:dalembertian_s}, \ref{eq:s_KJMA}, \ref{eq:rho_ini_KJMA}), 
the density of terminations is
\begin{equation}
	\rho_{\rm ter} (x,t) = 2v \left[ \int_{L_X^+} I \right] 
		\left[ \int_{L_X^-} I \right] s(x,t) \,.
\label{eq:rho_ter_KJMA}
\end{equation}
A termination at $X=(x,t)$ implies that no initiation occurs in $V_X$, one initiation occurs along $L_X^+$, and one along $L_X^-$.

\subsection{Rate equations for fork densities}

From the above formalism, we can easily recover the rate-equation formalism proposed in \cite{aGauthierPLoSOne2012} for fork densities. First, using Eq.~\eqref{eq:rho_pm}, the relation \eqref{eq:dalembertian_s} can be rewritten as
a rate equation for the density of right- or left-moving forks, 
\begin{equation}
	( \partial_t \pm v\partial_x ) \rho_\pm(x,t) 
		= \rho_{\rm init}(x,t) - \rho_{\rm ter}(x,t) \,. 
\end{equation}
Then, from Eqs.~(\ref{eq:rho_pm_KJMA}, \ref{eq:rho_ini_KJMA}, \ref{eq:rho_ter_KJMA}) we find \cite{aGauthierPLoSOne2012}, 
\begin{equation}
	( \partial_t \pm v\partial_x ) \rho_\pm(x,t)  
	= I s - 2v \,\frac{\rho_+ \rho_-}{s} \,. 
\end{equation}
Intuitively, fork densities change either because forks enter or leave a region (transport) or because there is initiation (birth) or termination (death).

\subsection{Correlations in replication timing}
\label{sec:correlations-reptiming}

As discussed in \cite{aYangMolSystBiol2010}, the observation that neighboring loci tend to have similar replication times can be fully consistent with the independent-firing assumption.  To more precisely quantify the correlation between replication times at different loci, we introduce the $N$-point unreplicated fraction $s(X_1,\cdots, X_N)$, where $X_i$ denotes the spacetime point $(x_i,t_i)$.  We define $s$ to be the fraction of cells where each of the $N$ loci $x_i$ is unreplicated at time $t_i$.  The joint probability distribution of replication timing at loci $x_1, \ldots , x_N$ is then
given by
 \begin{equation}
	P(X_1,\cdots,X_N) 
	= (-1)^N \partial_{t_1} \cdots \partial_{t_N} s(X_1,\cdots,X_N) \,. 
\end{equation}
In Fig.~\ref{fig:KJMA}(b) we note that each loci $x_i$ is unreplicated at time $t_i$
iff no initiations occur in $V_{X_1} \cup \cdots \cup V_{X_N}$, the union of past cones.  Therefore,
\begin{equation}
	s(X_1,\cdots,X_N)  = e^{-\int_{V_{X_1} \cup \cdots \cup V_{X_N}} dX' I(X')} \,. 
\label{eq:sN_corr}
\end{equation}
In \cite{aSekimotoPhysicaA1986}, Sekimoto derived an equivalent expression in the more-general setting of a time-dependent growth law.

To see why replication-fork propagation creates correlations between the replication times at different loci, consider the $N=2$ case.  Since $V_{X_1} \cup V_{X_2} = V_{X_1}+V_{X_2} - V_{X_1} \cap V_{X_2}$, the 2-point unreplicated fraction is equal to
\begin{equation}
	s(X_1,X_2) = s(X_1)s(X_2)e^{ \, + \int_{V_{X_1}\cap V_{X_2}} dX' \, I(X')} \,.
\label{eq:s_corr}
\end{equation} 
If the replication times at loci $x_1$ and $x_2$ were uncorrelated, both their probability distribution and their cumulative distribution would factor:  $P(X_1,X_2)=P(X_1)P(X_2)$ and $s(X_1,X_2)=s(X_1)s(X_2)$.  It is clear from Eq.~\eqref{eq:s_corr} that replication times at loci $x_1$ and $x_2$ are correlated because initiation events may occur in their common past cone $V_{X_1} \cap V_{X_2}$.  Indeed, if $I(X)$ is not identically zero in $V_{X_1}\cap V_{X_2}$, then $s(X_1,X_2) \neq s(X_1)s(X_2)$.  However, if the loci $x_1$ and $x_2$ are sufficiently far apart---that is, if $|x_1-x_2| \geq 2 v t_{\rm end}$, where $t_{\rm end}$ is the duration of S-phase---then their past cones do not intercept, and the replication times at $x_1$ and $x_2$ are indeed uncorrelated.

\subsection{The joint density of initiation}
\label{sec:correlations-init-density}

In Sec.~\ref{sec:correlations-reptiming}, we saw that the propagation of replication forks creates correlations in the timing of replication:  a location near an origin will tend to replicate soon after that origin fires.  A less obvious kind of correlation also exists in the initiation densities, where, again, we argue that apparent correlations can sometimes be deceptive.  Indeed, experimental observations of apparent origin synchrony \cite{aBerezneyChromosoma00} or of sequential firing, as observed in temporal transition regions \cite{aGuilbaudPLoSComputBiol2011}, suggest that  initiations may be temporally and spatially correlated, contradicting the independent-firing assumption.  Here, we will see that inferring independence from such observations can be subtle.  

In order to quantify the correlations observed in the distribution of initiations, we introduce the $N$-point joint density of initiations $\rho_{\rm init}(X_1,\cdots,X_N)$, defined as the probability to observe, during the same cell cycle, an initiation at each $X_i$.  Let us first assume that no $X_i$ belongs to the past cone of another $X_j$, as depicted in Fig.~\ref{fig:KJMA}(b).  Then, an initiation at each $X_i$ implies also that no initiation has occurred in $V_{X_1} \cup \cdots \cup V_{X_N}$.  Since the origins fire independently, the joint density of initiation is
\begin{equation}
	\rho_{\rm init}(X_1,\cdots, X_N) = I(X_1)\cdots I(X_N) s(X_1,\cdots,X_N) \,.
\label{eq:joint-init}
\end{equation}

To illustrate why replication fork propagation necessarily creates correlations in the joint density of initiation, we rewrite these expressions for $N=2$:
\begin{equation}
	\rho_{\rm init}(X_1,X_2) = \rho_{\rm init}(X_1)\rho_{\rm init}(X_2) 
		e^{ \, \int_{V_{X_1}\cap V_{X_2} } dX' I(X')} \,.
\label{eq:rho_ini_corr}
\end{equation} 
As in Eq.~\eqref{eq:s_corr}, initiation densities at $X_1$ and $X_2$ are correlated because of possible origin firing in their common past cone $V_{X_1} \cap V_{X_2}$.  To prove that neighboring initiations influence each other then takes more than the observation of initiation clusters or of sequential firing of nearby origins.  Only a clear departure from Eq.~(\ref{eq:rho_ini_corr}) would provide definitive evidence.  

Finally, if one of the $X_i$ belongs to the past cone of another $X_j$, $\rho_{\rm init}(X_1,\cdots ,X_N)$ is necessary null.  As re-replication is not allowed, we cannot observe an initiation in the future cone of another origin firing.  The joint density of initiation must satisfy this trivial correlation.

\subsection{Well-positioned replication origins}

In organisms such as the budding yeast \textit{S. cerevisiae}, origins initiate at predefined sites called \textit{potential origins}.  The local initiation rate then has the form \cite{aYangMolSystBiol2010}
\begin{equation}
	I(x,t)= \sum_{i} \delta(x-x_i) I_i(t) \,,
\label{eq:I_wellpos_oris}
\end{equation}
where $x_i$ is the position of potential origin $i$ and $I_i(t)$ its initiation rate.  All the analytical formulas derived in \cite{aYangMolSystBiol2010,aRetkutePhysRevE2012} are recovered as a particular case of the more general and compact expressions Eqs.~\eqref{eq:s_KJMA}--\eqref{eq:rho_ini_corr}, with the local initiation rate given by Eq.~\eqref{eq:I_wellpos_oris} %
\footnote{To make the connection with Refs. \cite{aYangMolSystBiol2010,aRetkutePhysRevE2012} more explicit, let us specify some of the quantities introduced in those references in terms of the local initiation rate.  The initiation probability density $\phi_i(t)$ in \cite{aYangMolSystBiol2010}, or origin activation time probability density $p_i(t)$ in \cite{aRetkutePhysRevE2012}, is given by $\phi_i(t) = p_i(t) = - \partial_t s_i(t) = I_i(t) s_i(t)$.  Note that this is not a normalized probability distribution, as $\int_0^{t_\infty} dt \, \phi_i(t) = q_i< 1$ is the potential efficiency/competence of origin $i$.  In \cite{aRetkutePhysRevE2012}, we also have $M_i(x,t) = s_i(t-|x-x_i|/v)$ and $p_i(x,t) = I_i(t-|x-x_i|/v) \, s_i(t-|x-x_i|/v)$.  In terms of the local initiation rate, the combinatorial expressions in \cite{aRetkutePhysRevE2012} simplify greatly; for instance, $p_i(x,t) \prod_{j\neq i} M_j(x,t)=I_i(t-|x-x_i|/v) \, s(x,t)$.}.

Let us specify the expressions for $s(x,t)$ and $\rho_{\rm init}(x)$ in the case of well-positioned origins. From Eqs.~\eqref{eq:s_KJMA} and \eqref{eq:I_wellpos_oris}, the unreplicated fraction can be written
\begin{align}
	s(x,t) &= \prod_i s_i\left(t-\frac{|x-x_i|}{v} \right) \, , \\
	\mathrm{where} \quad s_i(t) &\equiv e^{-\int_0^t dt' \, I_i(t')}
	\label{eq:si-def}
\end{align} 
is the probability that the potential origin $i$ has yet not initiated at time $t$. In words, the locus $x$ is unreplicated a time $t$ iff each origin $i$ has not initiated before time $t-|x-x_i|/v$. From  Eqs.~\eqref{eq:rho_ini_KJMA} and \eqref{eq:I_wellpos_oris}, the initiation density profile  will have sharp peaks at potential-origin sites:
\begin{align}
	\rho_{\rm init}(x) &= \sum_{i} \delta(x-x_i) E_i \,, \nonumber \\
	\mathrm{with} \quad E_i &= \int_0^{t_\infty} dt'\, I_i(t') s(x_i,t') \,, 
\label{eq:efficiency}
\end{align}
where $E_i$, the \textit{observed efficiency} of origin $i$, is defined as the fraction of cells where the origin $i$ has activated before the end of S phase. The observed efficiency of the origin $i$ depends on its initiation properties but is also affected by the initiation properties of neighboring origins \cite{aYangMolSystBiol2010,aRetkutePhysRevE2012}. Indeed, when the locus $x_i$ is replicated by a fork coming from a neighboring origin, the potential origin $i$ will not be activated during this cell cycle, and the potential origin is \textit{passively replicated}. It is then interesting to consider the \textit{potential efficiency} of a replication origin---the probability that the origin would activate during S-phase if passive replication by neighboring origins is prevented.  The potential efficiency $q_i$ of origin~$i$, denoted \textit{origin competence} in \cite{aMouraNucleicAcidsRes2010,aRetkutePhysRevE2012}, is equal to
\begin{equation}
	q_i = 1- e^{-\int_0^{t_\infty} dt' \, I_i(t')} = 1-s_i(t_\infty) \,,  
\label{eq:def_competence}
\end{equation}
as $s_i(t)$ is the probability that the origin $i$ has not yet initiated at time $t$. Contrary to a claim in \cite{aRetkutePhysRevE2012}, the KJMA formalism does not assume $100\%$ competent origins; in general, $q_i<1$\footnote{For example, we can have $q_i<1$ if origins fail to be licensed prior to the start of S phase \cite{aMouraNucleicAcidsRes2010,aRetkutePhysRevE2012}.  Let the  \textit{licensing probability} for origin $i$ be $L_i$.  Then $s_i(t) = (1-L_i)+L_i e^{-I't}$, where $I'$ is the initiation rate \textit{if licensed}, assumed, for simplicity to be constant for all origins and all time.  From Eq.~\eqref{eq:si-def}, $I_i(t) = -\tfrac{d}{dt} \ln s_i(t) \sim I' \bigl( \tfrac{L_i}{1-L_i} \bigr) e^{-I' t}$ for times $t \gg 1/I'$.  A finite licensing probability thus cuts off the effective initiation rate at long times, and the failure to license origins can be absorbed into the effective initiation rate.  Note that if $L_i=1$, we recover $I_i = I'$.}.  In budding yeast, passive replication has a strong impact on the efficiencies of replication origins:  the observed efficiency is usually much smaller than the potential efficiency \cite{aYangMolSystBiol2010}.

\section{Example replication program}
\label{sec:example-rep-prog}

\begin{figure}[hbt]
  \includegraphics[width=1.0\linewidth]{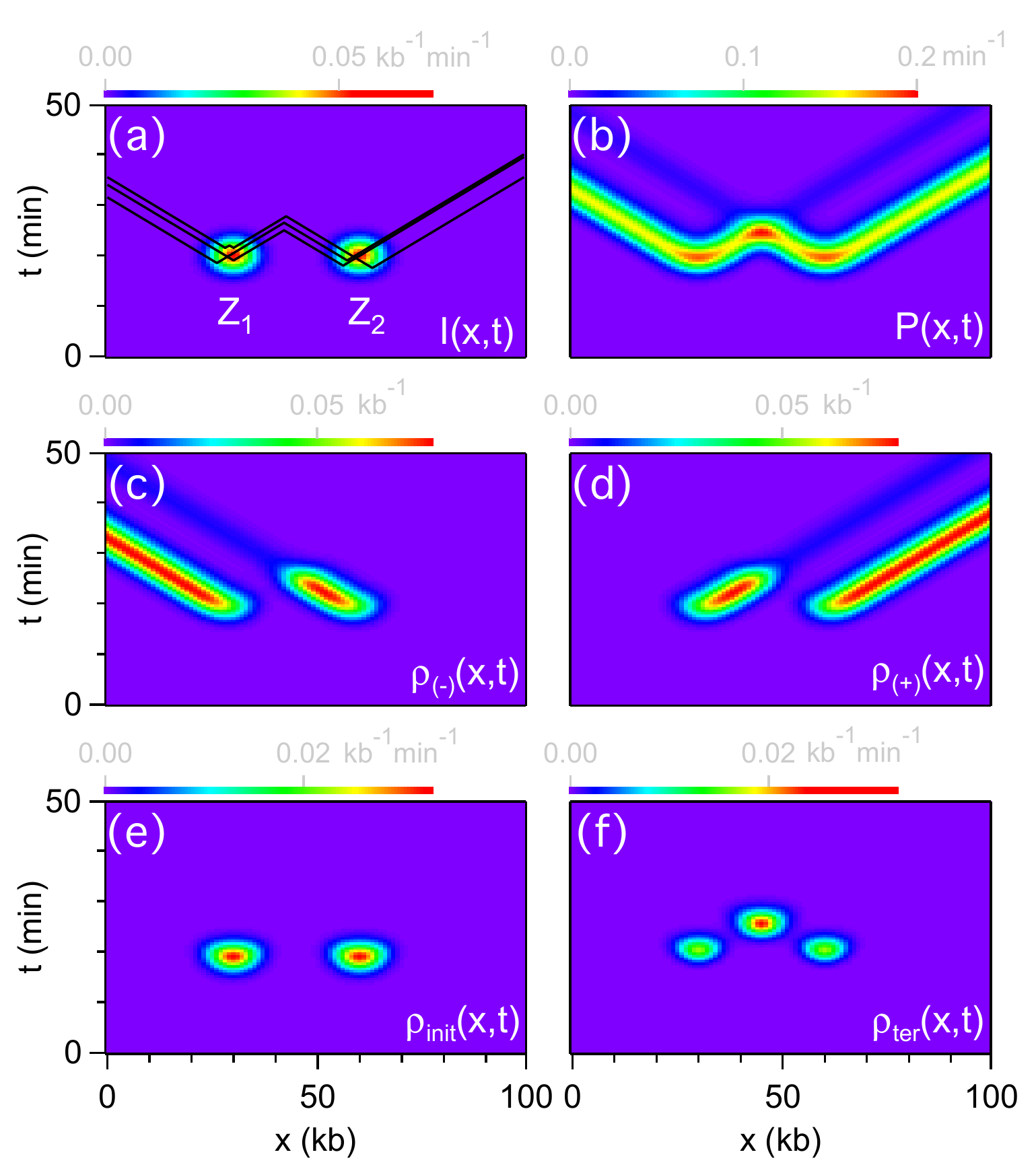}
  \caption{
    \label{fig:gaussian_blobs}
    (Color online) 
    Replication program with two extended initiation zones $Z_1$ and $Z_2$.
    (a) Heatmap of the local initiation rate $I(x,t)$.
    The black lines correspond to single cell cycle realisation of the replication program,
    obtained by Monte Carlo simulation.
    (b) Replication distribution, Eq.~\eqref{eq:P_KJMA}.
    (c,d) Densities of left- and right-moving forks, Eq.~\eqref{eq:rho_pm_KJMA}.
    (e,f) Densities of initiation, (Eq.~\ref{eq:rho_ini_KJMA}) and termination 
    	(Eq.~\ref{eq:rho_ter_KJMA}). 
  }
\end{figure}

Let us now illustrate the formalism developed in the two preceding sections on an  artificial replication program that consists of two extended initiation zones, $Z_1$ and $Z_2$.  In Fig.~\ref{fig:gaussian_blobs}(a), the spacetime representation of the local initiation rate is color coded by a heat map.  To give an idea of the resulting stochasticity, we sample by Monte Carlo simulation five realizations of the replication program, represented by the black lines on Figs.~\ref{fig:gaussian_blobs}(a).  Several aspects of the replication program, analytically derived from the local initiation rate using the results of Sections~\ref{sec:gen-rep} and \ref{sec:indep-ori-firing}, are represented on Figs.~\ref{fig:gaussian_blobs}(b)-(f).

Notice how Fig.~\ref{fig:gaussian_blobs} reveals many fine details about the replication process.  For example, the density of termination events in Fig.~\ref{fig:gaussian_blobs}(f) shows three zones.  At the center is the strongest one, representing the case where forks from the two origin regions collide after propagating roughly to the midpoint between the initiation zones $Z_1$ and $Z_2$.  The two weaker termination zones overlap with the initiation zones and represent cases where two or more initiation events within the same zone lead to a fork collision soon after the initiation event.  Solving the analytical model allows us to detect and quantify the probability for these different scenarios to occur.

\section{Inferring the local initiation rate}
\label{sec:inferring-local-init-rate}

In Sections~\ref{sec:gen-rep}--\ref{sec:example-rep-prog}, we showed how to solve the \textit{forward} problem of replication: given an initiation rate $I(x,t)$, calculate various quantities of interest for the replication process, for example the unreplicated fraction $s(x,t)$.  Now we consider the \textit{inverse} problem:  given a noisy measurement of $s(x,t)$, can we infer $I(x,t)$?  In particular, we advance a new, non-parametric method that avoids having to define a model structure for $I(x,t)$.

To test the new method under well-controlled circumstances, we will focus on inverting simulated data based on the  spatiotemporal replication program presented in Section~\ref{sec:example-rep-prog}.  The data will have a spacetime resolution comparable to that of present experiments and will include noise levels that are also typical.

We begin by first reviewing past attempts to solve this inverse problem, including fitting strategies and analytic approaches based on expressing the initiation rate $I(x,t)$ as a function of the non-replicated fraction $s(x,t)$.  After discussing the limitations of previous attempts, we then propose a Bayesian, non-parametric approach to infer $I(x,t)$ from replication timing data.  We will test this inference scheme on the artificial data set described above and show that near-perfect reconstruction of the replication program (with negligible posterior uncertainty) is attained for many quantities of interest, such as the unreplicated fraction, the densities of replication forks, the densities of initiation and termination. The local initiation rate is also inferred with low posterior uncertainty in most regions except at the end of S-phase, where the unreplicated fraction, already close to zero, is insensitive to large variations in the initiation rate.

\subsection{Curve-fitting strategies}
\label{sec:curve-fitting}

As discussed in the Introduction (Sec.~\ref{sec:intro}), the replication fork velocity $v$ and initiation function $I(x,t)$ can be estimated by curve fitting \cite{herrick02,aMouraNucleicAcidsRes2010,aLuoBMCBioinformatics2010,aYangMolSystBiol2010,hawkins13,demczuk12}.  The main issue is that one must make strong assumptions about the prior functional form for $I(x,t)$, for example whether origins are localized along the genome, the type of time dependence, etc.  Besides requiring \textit{a priori} knowledge about the biology that is not always available, the underlying forms may not really be what is assumed.  Also, the number of parameters needed is not clear in advance.  For example, the number of detectable origins in budding yeast  is an output of the inference process.   In addition, one needs to provide initial values for all parameters.

For all these reasons, a successful curve fit requires both \textit{a priori} knowledge and a good level of technical expertise.  Below, we will explore a strategy that requires only vague \textit{a priori} expectations and that can, in principle, be automated.

\subsection{Exact inverse}
\label{sec:exact-inverse}

Recently, we showed how to invert explicitly the KJMA formula Eq.~\eqref{eq:s_KJMA}, thereby determining analytically $I(x,t)$ from $s(x,t)$ \cite{aBakerPhysRevLett2012} \footnote{
Generalizing Eq.~\eqref{eq:I_KJMA} to a space- and time-dependent velocity field $v(x,t)$ is straightforward, albeit cumbersome \cite{baker11}:  $I(x,t) = \tfrac{1}{2} \left[ \left( \tfrac{1}{v} \partial_t v \right) \tfrac{1}{v} \partial_t + (\partial_x v) \partial_x - v \square \right] \ln s(x,t)$.  Because it is not at present clear whether systematic (as opposed to random) variation of fork velocities is important, we focus on the constant-$v$ case.
}:
\begin{equation}
	I(x,t) = -\tfrac{1}{2} v \square \ln s(x,t) \,.
\label{eq:I_KJMA}
\end{equation}
Because Eq.~\eqref{eq:I_KJMA} gives an exact expression for $I(x,t)$, it would seem to provide an alternative to curve-fit approaches: rather than guess the form of $I(x,t)$, we can simply calculate it from the data, $s(x,t)$.  Unfortunately, the analytical inverse is numerically unstable: taking two derivatives amplifies noise tremendously.  Thus, Eq.~\eqref{eq:I_KJMA} can be naively applied only if essentially noise-free data for $s(x,t)$ are available.  For example, in \cite{aBakerPhysRevLett2012}, we used Eq.~\eqref{eq:I_KJMA} to invert simulations that had negligible numerical noise.  When applied directly to low-resolution experimental data with realistic amounts of noise, Eq.~\eqref{eq:I_KJMA} gives unphysical results such as negative initiation rates \cite{herrick02}.  Simple \textit{ad hoc} fixes, such as smoothing $s(x,t)$ over fixed space and time scales \cite{aBakerPhysRevLett2012}, lead to unacceptable distortion in the estimate of $I(x,t)$ and also do not give uncertainties in estimated initiation rates.  All of these shortcomings motivate a more fundamental approach.

\subsection{Bayesian inference}
\label{sec:Bayes}

Here, we will adopt a Bayesian, non-parametric approach to more properly infer $I(x,t)$ from replication timing data.  Bayesian methods offer a consistent and conceptually well-founded framework for inference, where all assumptions are explicitly stated \cite{Jaynes2003}.

\subsubsection{Introduction}

The Bayesian formulation is well adapted to parameter-estimation problems \cite{Jaynes2003}. In our case, the goal is to infer the parameter $I$ (the local initiation rate) from the data $d$ (a noisy measurement of the unreplicated fraction).  We recall that the posterior probability of $I$, given data $d$, is determined by Bayes' theorem, which is derived from the product and sum rules of probability theory \cite{Jaynes2003}:
\begin{equation}
	\underbrace{P(I|d,\beta)}_{\mathrm{posterior}} = \frac{1}{P(d|\beta)} 
		\underbrace{P(d|I,\beta)}_{\mathrm{likelihood}} 
		\underbrace{P(I|\beta)}_{\mathrm{prior}} \,, 
\label{eq:Bayes}
\end{equation}
where the normalizing factor, the \textit{evidence}, is given by
\begin{equation}
	\underbrace{P(d|\beta)}_{\mathrm{evidence}} 
		= \int dI \, P(d|I,\beta) \, P(I|\beta) \,. 
\label{eq:evidence}
\end{equation}
In Eq.~\eqref{eq:Bayes}, the likelihood follows the noise model for the data, while the prior encodes any available information---even vague---about the parameter to infer; in replication, for instance, we know that initiation rates $I(x,t)$ must be positive.  We also expect that temporal and spatial variations of $I(x,t)$ are smooth, although we may not know the smoothness scales.  Below, we will describe in more detail the probabilistic model used for inference given such vague priors.

Often, the specification of a probabilistic model for the likelihood and the prior requires an additional set of parameters, called \textit{hyperparameters}, symbolized by $\beta$ in Eqs.~\eqref{eq:Bayes} and \eqref{eq:evidence}.  In our case, the hyperparameters comprise the fork velocity $v$, which affects the relationship Eq.~\eqref{eq:s_KJMA} between the unreplicated fraction data and the initiation rate, the noise level affecting the data, and additional parameters encoding prior information about the initiation rate, for example the temporal scale of smoothness.  These  hyperparameters can themselves be inferred by another application of Bayes' theorem  \cite{MacKay2003}:
\begin{equation}
	P(\beta | d) 
	= \frac{1}{P(d)} P(d | \beta) P(\beta) \,.
\label{eq:hyperparam}
\end{equation}
The posterior probability of the hyperparameters is thus proportional to the evidence and the prior probability of the hyperparameters.  Given the posterior $P(\beta | d)$, we can eliminate the hyperparameters by \textit{marginalization}, or ``integrating out."  For example,
\begin{align}
	P(I|d) = \int P(I|d,\beta) \, P(\beta | d) \, d\beta \,.
\end{align}

The Bayesian formulation is also well adapted to model selection.  Given data and candidate theories, Bayes' theorem allows one to estimate the most probable model  \cite{MacKay2003}.  For instance, we could compare the probabilistic model presented here and the fitting procedure (that can easily be reformulated in a Bayesian framework) employed in yeast.  We could even compare to a theoretical model that extends the KJMA formalism to take into account correlations in the origin firing.  Such model comparisons are beyond the scope of the present paper.

The inference task here is complicated by the nonlinear relationship Eq.~(\ref{eq:s_KJMA}) between the data (the unreplicated fraction profiles) and the initiation rate we seek to infer and by the positivity constraint on the initiation rate.  Indeed, if the relationship were linear and no positivity constraint needed to be enforced, then we would be able to derive the posterior Eq.~\eqref{eq:Bayes} analytically.  Below, we will approximate the posterior probability distribution by its mode, the \textit{maximum a posteriori} (MAP) approximation, which requires a high-dimensional nonlinear optimization algorithm. To estimate the width of the posterior, we will sample directly the posterior by Markov chain Monte Carlo (MCMC) techniques. Finally, to estimate the evidence, we will use the \textit{Laplace approximation}, which is the analog of the saddle-point approximation in statistical physics.

\subsubsection{Likelihood}

We model the data as a noisy version of the unreplicated fraction $s$, sampled in time and space: 
\begin{align}
	d_k  &= s(x_k,t_k) + \xi_k ,  \nonumber \\[3pt]
	\mathrm{with} \quad s(x,t) &= e^{-\int_{V_X} dx' \, dt' \, I(x',t')} \,,
\label{eq:data}
\end{align}
with noise described by independent, identically distributed (i.i.d.) Gaussian random variables of standard deviation $\sigma_d$.  Thus, $\xi_k \sim \mathcal{N}(0,\sigma_d^2)$, and the likelihood is
\begin{align}
	P(d | I, v, \sigma_d) &= P_{\mathrm{noise}} (d-s) \nonumber \\
	&= \prod_{k} \frac{1}{\sqrt{2 \pi \sigma_d^2}} \, e^{-\frac{1}{2\sigma_d^2} 
		[d_k-s(x_k,t_k)]^2} \,,
\label{eq:likelihood}
\end{align}
where the product is over all data points $k$.

\begin{figure}[hbt]
	\includegraphics[width=0.95\linewidth]{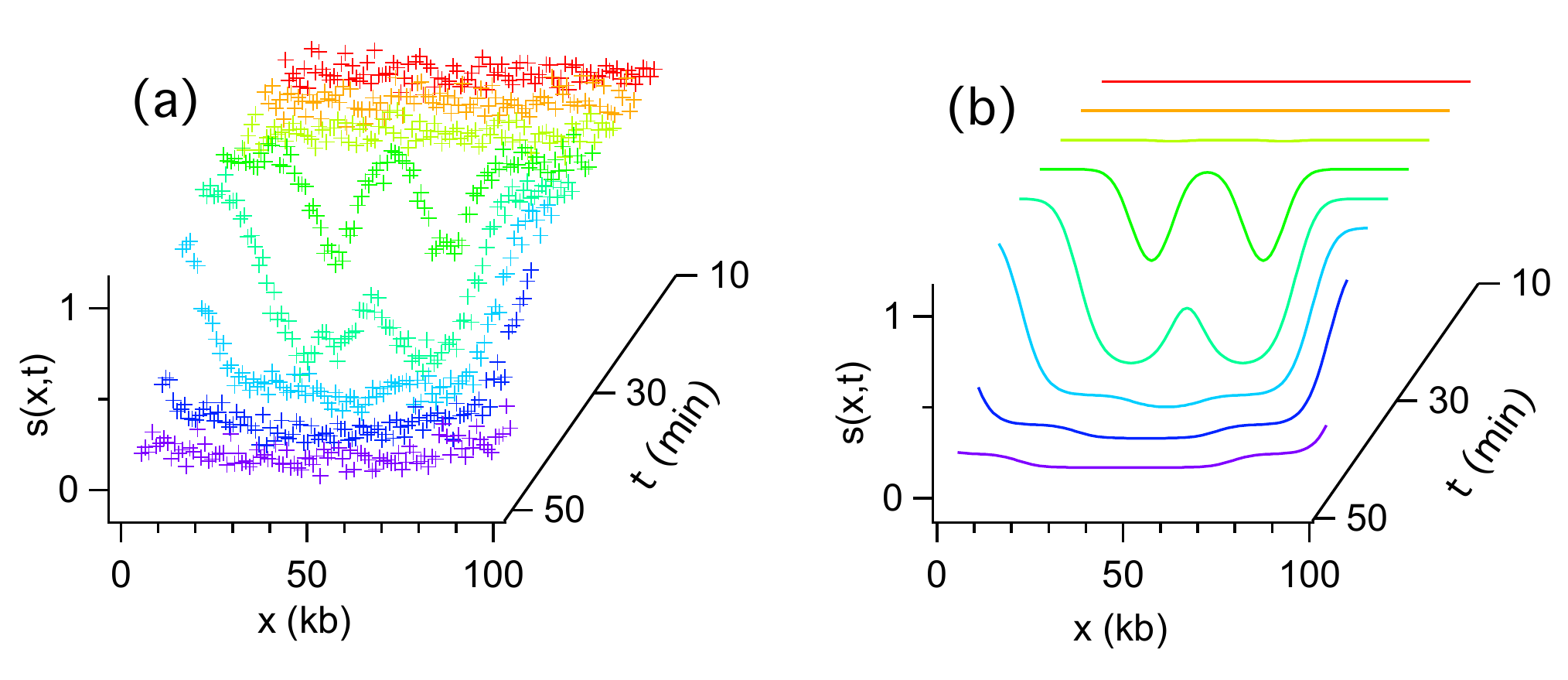}
	\caption{(Color online) Simulation of the replication program with extended initiation zones (Fig.~\ref{fig:gaussian_blobs}).  (a) Artificial data set generated by adding Gaussian noise of standard deviation $\sigma_d=0.05$ to the true unreplicated fractions in (b).  In (a) and (b), the unreplicated fraction profiles are given every 1 kb and every 5 min from $t=10 $ min (red) to $t = 50$ min (purple).
}
\label{fig:artificial_data}
\end{figure}

In the artificial data set shown in Fig.~\ref{fig:artificial_data}, the noisy unreplicated fractions are sampled every 1 kb in a fragment of 100 kb and every 5 min from $t=10$ min to $t = 50$ min.  These resolutions match that of the recent budding-yeast experiments described above.  We chose $\sigma_d = 0.05$, again typical of current experiments \cite{aYangMolSystBiol2010, muller13}.  Note that Fig.~\ref{fig:artificial_data} can also be interpreted as a plot of replicated fraction $f=1-s$ from times of 10 to 50 min.

\paragraph*{Comment on the noise model.}

Although we model the noise by i.i.d. Gaussian random variables of standard deviation $\sigma_d$, it is straightforward to substitute any noise model in Eq.~(\ref{eq:likelihood}), including correlations, time- or space-dependent variance, or non-Gaussian distributions.  As a real-world example, the analysis of data on budding yeast showed a variance that increased throughout S phase and a noise distribution,  that while Gaussian for small fluctuations, was exponential for larger ones \cite{aYangMolSystBiol2010}.  In general, small deviations from the Gaussian form will not affect the analysis much.

\subsubsection{Prior}

A key advantage of the Bayesian formulation is that we can specify the prior, the set of possible initiation rate functions, without having to impose a particular functional form.  Nevertheless, we do have some vague prior knowledge about $I(x,t)$ that should be used to constrain the set of possible initiation functions: it must be positive and its temporal variations are smooth. In some cases, spatial variations are also smooth.

To ensure the positivity of the initiation rate, we change variables, defining 
\begin{equation}
	I(x,t) \equiv I_0 \, 10^{m(x,t)} \,.
\end{equation}
In other words, rather than trying to infer the initiation rate $I$ directly, we will infer its logarithm, $m$.  

To enforce smooth variations in the initiation rate, we will use a Gaussian process prior \cite{Rasmussen2006} on $m = \log_{10}(I/I_0)$:
\begin{equation}
	m \sim \mathcal{GP}(0,\Sigma) \,,
\label{eq:GP}
\end{equation}
with a homogeneous, squared-exponential covariance function that depends on the spatial separation $\Delta x$ and the temporal separation $\Delta t$: 
\begin{equation}
	\Sigma(x,t; \, x+\Delta x,t+\Delta t) = \sigma_0^2 \, 
		e^{-\left( \frac{\Delta x}{\ell_0} \right)^2} 
		e^{-\left( \frac{\Delta t}{\tau_0} \right)^2} \,.
\label{eq:Sigma}
\end{equation}

A Gaussian process $m$ can be viewed as the infinite-dimensional analog of the multivariate normal distribution; it defines a probability distribution over functions.  The precise definition is that the values of $m$ at an arbitrary set of points $(X_1,\ldots , X_N)$ are distributed according to the multivariate normal distribution $[m(X_1),\ldots,m(X_N)] \sim \mathcal{N}(0,\Sigma)$, with covariance matrix  $\Sigma_{ij} = \Sigma(X_i,X_j)$.  In our case, we would like to infer the initiation rate at a spatial resolution of $\delta x= 1$ kb and a temporal resolution of  $\delta t= 0.5$ min (we set $\delta t$ in order to have  $ \delta x = v \, \delta t$, with a fork velocity equal to $v=2$ kb.min$^{-1}$).  This defines the grid of points $X \equiv (x,t)$ where $m$ should be evaluated. The prior distribution on $m = \{m(x,t)\}$ is therefore the multivariate normal
\begin{equation}
\label{eq:prior}
	P(m | \sigma_0, \tau_0 ,\ell_0 ) 
	= \frac{1}{\sqrt{\det (2\pi\Sigma )}} e^{-\frac{1}{2}m \cdot \Sigma^{-1} m} \,,
\end{equation}
with covariance function $\Sigma$ evaluated at the grid of points $(x,t)$ using Eq.~\eqref{eq:Sigma}.  In the covariance function Eq.~\eqref{eq:Sigma}, $\sigma_0$ quantifies the prior expectations about the range of values taken by $m$. The square-exponential decay as a function of the time interval separating two points, on a characteristic time scale $\tau_0$, enforces the smoothness of the function $m$ on the same time scale, and similarly for the spatial scale $\ell_0$.  The limit $\ell_0 \rightarrow 0$ means that $m$ values at different genomic positions are uncorrelated. It is obtained by replacing the squared exponential in Eq.~\eqref{eq:Sigma} by a Dirac delta function, $\delta(\Delta x)$.  

In Gaussian process regression, the task is to go from a Gaussian-prior representation of $m(x,t)$ (Eq.~\ref{eq:GP}) to a posterior representation that incorporates the noisy observations $d_k$.  Note that many authors define a Gaussian process regression to be one where the posterior distribution for $m$ is also a Gaussian process (that is, they assume that the data are related to $m$ by a linear transformation).  Here, the data and $m$ are nonlinearly related, and the resulting distribution for $m$ is non-Gaussian.  For simplicity, we also refer to this case as Gaussian process regression, but we will need to use special techniques to deal with the non-Gaussian nature of the posterior distribution.

\subsubsection{Hyperparameters}

As discussed earlier, we can estimate the hyperparameters from the data set itself.  Here, instead of carrying out this procedure for all of them, we will do so only for the most interesting ones, the fork velocity $v$ and the spatial scale $\ell_0$ for $I(x,t)$ variations.  The latter is especially delicate, in that some organisms, such as budding yeast, have near $\delta$-function initiation sites, while others, such as frog embryos, permit initiation anywhere and have slowly varying densities.  Accordingly, we will carry out the self-consistent selection for these parameters below.

We first fix the hyperparameters of lesser interest.  For example, $\sigma_0$ and $I_0$ set the range of values allowed for the initiation rate.  Their precise value should not matter much, as long as the allowed range of values is larger than the actual range of values taken by the initiation rate.  Here, we choose $I_0=10^{-4}$ kb$^{-1}$ min$^{-1}$ and $\sigma_0=3$, to allow for a very wide range of values for the initiation rate.  This choice allows a 1-$\sigma$ range of initiation rates of between 10$^{-1}$ and 10$^{-7}$ kb$^{-1}$ min$^{-1}$.  

The temporal scale $\tau_0$ defines how quickly $I(x,t)$ can vary.  Although in principle as interesting as the spatial scale $\ell_0$, the evidence to date suggests that the experimental range of values is much narrower.  For example, previous analysis of the replication kinetics in yeast \cite{aYangMolSystBiol2010} is consistent with $\tau_0 \approx 10$ min., about $1/4$ the duration of S phase, and we used this value in the inference procedure.  

\begin{figure}[hbt]
	\includegraphics[width=0.95\linewidth]{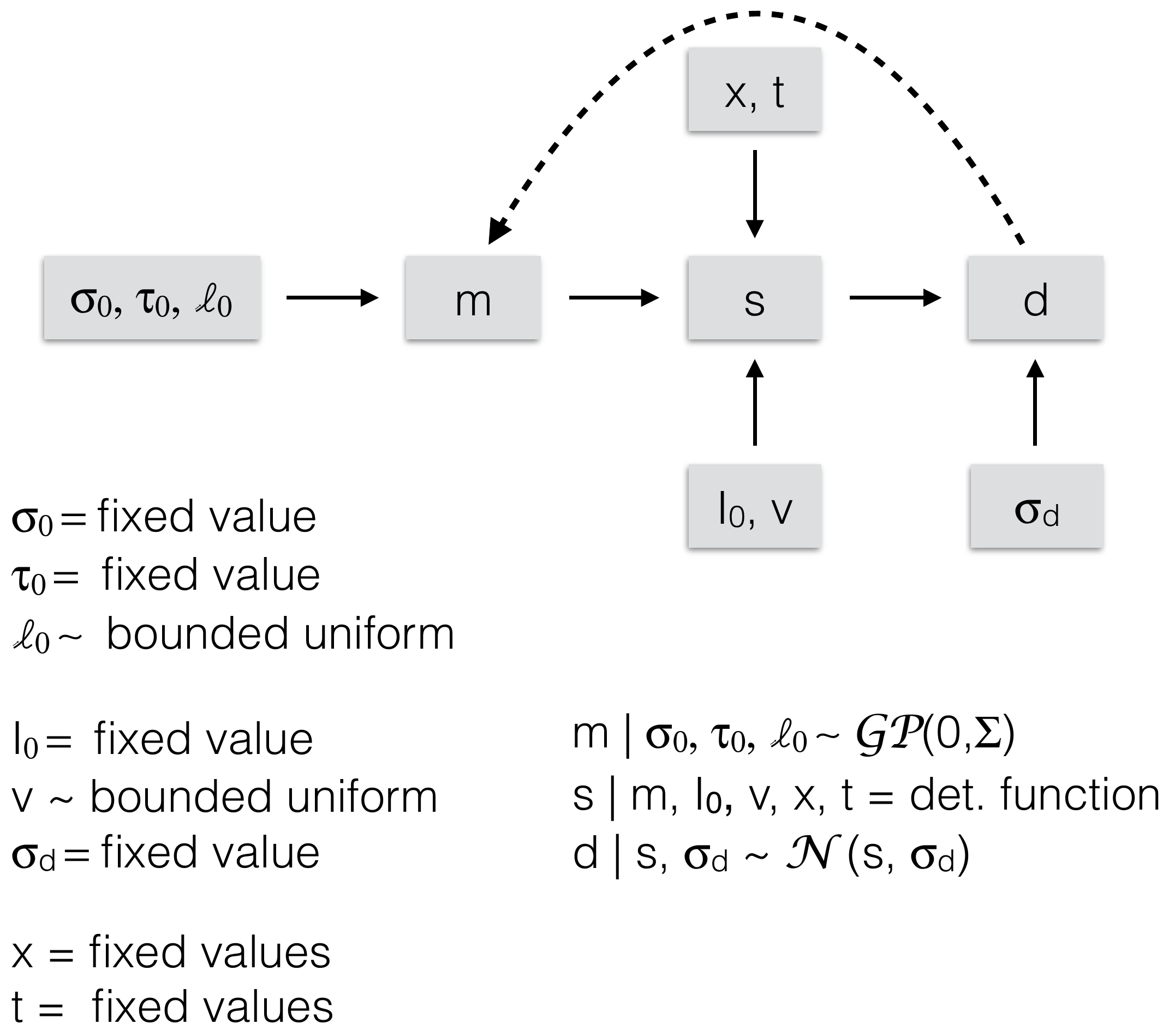}
	\caption{Diagram summarizing the forward replication model $m \to d$ evaluated at grid points ($x$, $t$) and its hyperparameters ($\sigma_0$,~$\tau_0$,~$\ell_0$,~$I_0$, $v$, $\sigma_d$).  The symbol ``$\sim$" means ``distributed as," and the dashed arrow denotes the inference $d \to m$.}
\label{fig:diagram}
\end{figure}

The complete probabilistic model is summarized in Fig.~\ref{fig:diagram}.  Below, we  first use the model to infer the log initiation rate $m = \log(I/I_0)$ from the data $d$, assuming the hyperparameters to be known.  In the last subsection, we will solve for $m_{\rm MAP}$ over a grid of values for $v$ and $\ell_0$ and find that the posterior is almost entirely concentrated at the correct (simulation) values.

\subsubsection{Posterior}

The posterior $P( m | d,  \beta)$ for the log initiation rate $m = \log_{10}(I/I_0)$ is given by Bayes' theorem Eq.~(\ref{eq:Bayes}), with the likelihood given by Eq.~\eqref{eq:likelihood}, the prior given by Eq.~\eqref{eq:prior}, and the hyperparameters $\beta = \{ v, \sigma_d, I_0, \sigma_0, \tau_0 , \ell_0 \}$.  Note that the parameter $m$ to infer is evaluated at a resolution of 1 kb in space and 0.5 min in time and thus forms an $N_x N_t = 100 \times 100$ dimensional vector.  Thus, the posterior for $m$ is a probability distribution defined on a very high dimensional ($10^4$) space.  Below, we will consider both replacing the distribution by its mode (maximum \textit{a posteriori} approximation) and sampling the posterior by MCMC techniques.

\subsubsection{Maximum \textit{a posteriori} approximation}

The mode of the posterior distribution, which gives the \textit{maximum a posteriori} (MAP) estimate, can be found by minimizing the ``energy'' functional \cite{lemm03,bialek12}
\begin{align}
	E(m) &= - \ln P( m,d | \beta) \nonumber \\
	&= - \ln P(d | m,\beta) - \ln P( m | \beta) \nonumber \\
  	&= \tfrac{1}{2\sigma_d^2} \sum_k [d_k-s(x_k,t_k)]^2 
		+ \tfrac{1}{2} N_d \ln 2\pi \sigma_d^2 \nonumber \\
	&+  \tfrac{1}{2}m \cdot \Sigma^{-1} m + \tfrac{1}{2} \ln \det (2\pi \Sigma) 
		\nonumber \\ 
	\mathrm{with} \quad s(x,t) &= 
		e^{-\int_{V_X} dx' \, dt' \, I_0 10^{m(x',t')}} \,.
\label{eq:energy}
\end{align}
The quantity $E(m)$ is the negative log of the joint posterior, with $N_d$ the number of data points. The MAP estimate, $m_{\rm MAP} = \argmin E(m)$, can be interpreted as a compromise between minimizing the least-square fit $\frac{1}{2\sigma_d^2} (d-s)^2$ (the ``energy") and minimizing $\frac{1}{2}m \cdot \Sigma^{-1} m$ (the ``entropy"), where smoother states are lower entropy because they are compatible with fewer data sets. Alternatively, we can view the minimization as a regularized ``Tikhonov" inverse \cite{ibJaynes1984}, where the compromise is between finding the $m$ that best reproduces the data $d$ and minimizing the Tikhonov penalty term, which favors smooth $m$ on the spatial scale $\ell_0$ and temporal scale $\tau_0$. 

We minimized $E$ in Eq.~\eqref{eq:energy} numerically via the Newton conjugate gradient algorithm \cite{Nocedal2006}.  Although we minimize in a $10^4$-dimensional space, the program converges in less than a minute on a regular laptop.

\begin{figure}[hbt]
  \includegraphics[width=1.0\linewidth]{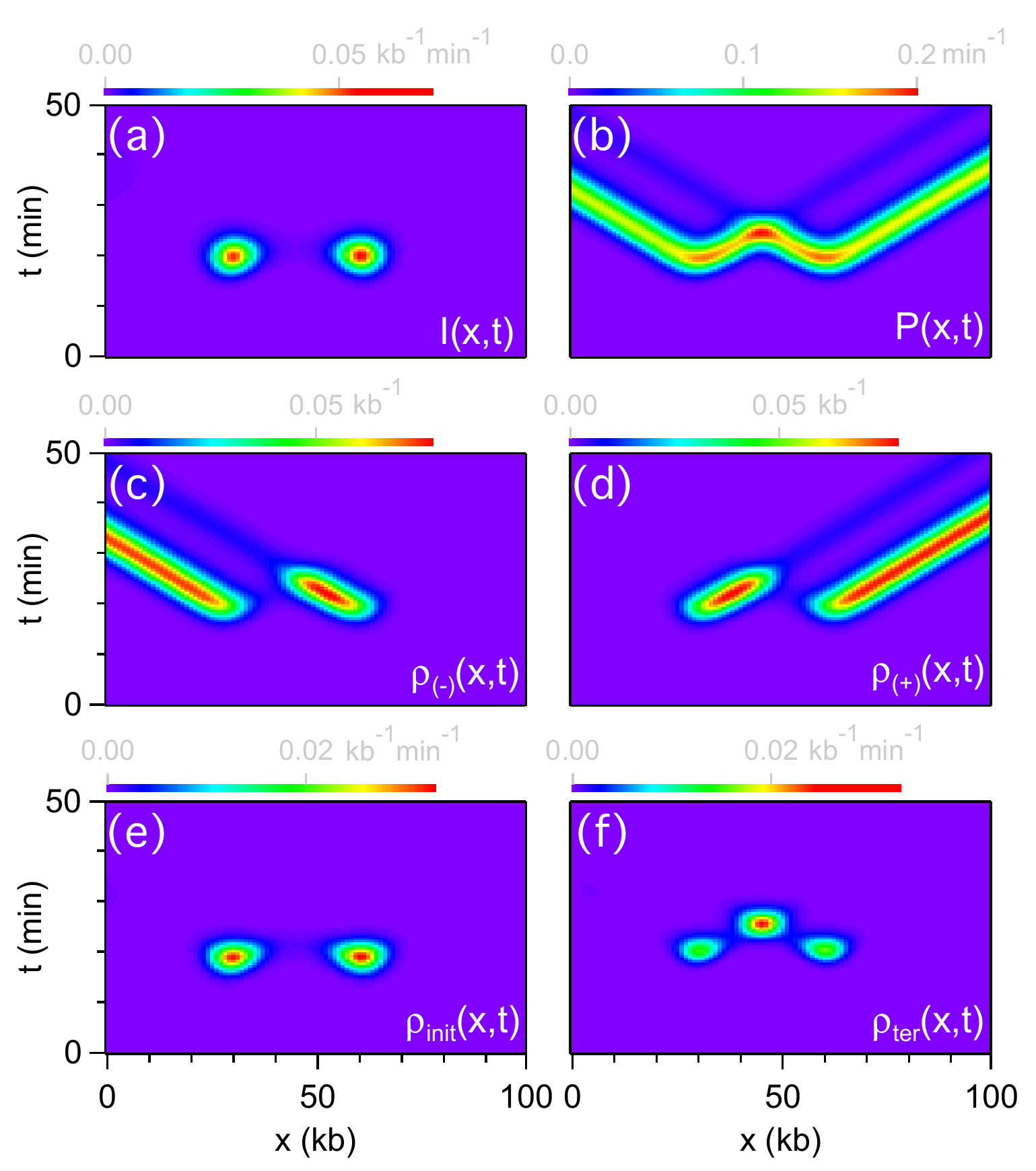}
  \caption{
    \label{fig:reconstruction}
    (Color online) 
    Near-perfect reconstruction of the replication program in Fig.~\ref{fig:gaussian_blobs}.
    All characteristics of the replication program are reconstructed using
    the MAP estimate $m_{\rm MAP}$ of $m = \log_{10}(I/I_0)$. 
    (a) Local initiation rate $I_{\rm MAP}= I_0 \, 10^{m_{\rm MAP}}$. 
    (b) Replication distribution, Eq.~\eqref{eq:P_KJMA}.
    (c,d) Densities of left- and right-moving forks, Eq.~\eqref{eq:rho_pm_KJMA}.
    (e,f) Densities of initiation, (Eq.~\ref{eq:rho_ini_KJMA}) and termination 
    	(Eq.~\ref{eq:rho_ter_KJMA}). 
  }
\end{figure}

The MAP approximation is to replace the posterior distribution by a Dirac $\delta$-function at its mode, 
\begin{equation}
	P( m | d,\beta) \simeq \delta(m-m_{\rm MAP}) \,.
\label{eq:MAP_approx}
\end{equation}
That is, we simply substitute $m_{\rm MAP}$ into the analytical expression of the initiation rate and into all other quantities of interest.  As shown in Fig.~\ref{fig:reconstruction}(a), the estimated local initiation rate $I_{\rm MAP} = I_0 10^{m_{\rm MAP}}$ is very close to the true initiation rate, Fig.~\ref{fig:gaussian_blobs}(a).  Similarly, the estimated unreplicated fraction, Fig.~\ref{fig:reconstruction}(b), density of right- and left-moving forks, Fig.~\ref{fig:reconstruction}(c)--(d), as well as the density of initiation, Fig.~\ref{fig:reconstruction}(e), and termination, Fig.~\ref{fig:reconstruction}(f), obtained by simply substituting $I_{\rm MAP}$ in the analytical expressions of Section~\ref{sec:indep-ori-firing} are indistinguishable from their true values in Fig.~\ref{fig:gaussian_blobs}.  Finally, note that all those quantities are reconstructed at the desired temporal resolution of 0.5 min, while the original data $d$ in Fig.~\ref{fig:artificial_data} has only a temporal resolution of 5 min. This interpolation is possible because the temporal smoothness scale $\tau_0=10$ min.

\subsubsection{MCMC sampling of the posterior}

The MAP approximation Eq.~(\ref{eq:MAP_approx}) would seem to be a rather crude one, as it neglects the posterior uncertainty for $m$.  Moreover, the MAP estimate $m_{\rm MAP}$ is usually not a representative sample from the posterior, and its value is not invariant under re-parametrization \cite{MacKay2003}.  However, in our particular case, the MAP estimate $m_{\rm MAP}$ does yield a very accurate reconstruction of the replication program:  Since, as we will see below, the posterior uncertainty for most quantities turns out to be negligible, samples from the posterior distribution are almost always close to the MAP value.

\begin{figure}[hbt]
  \includegraphics[width=1.0\linewidth]{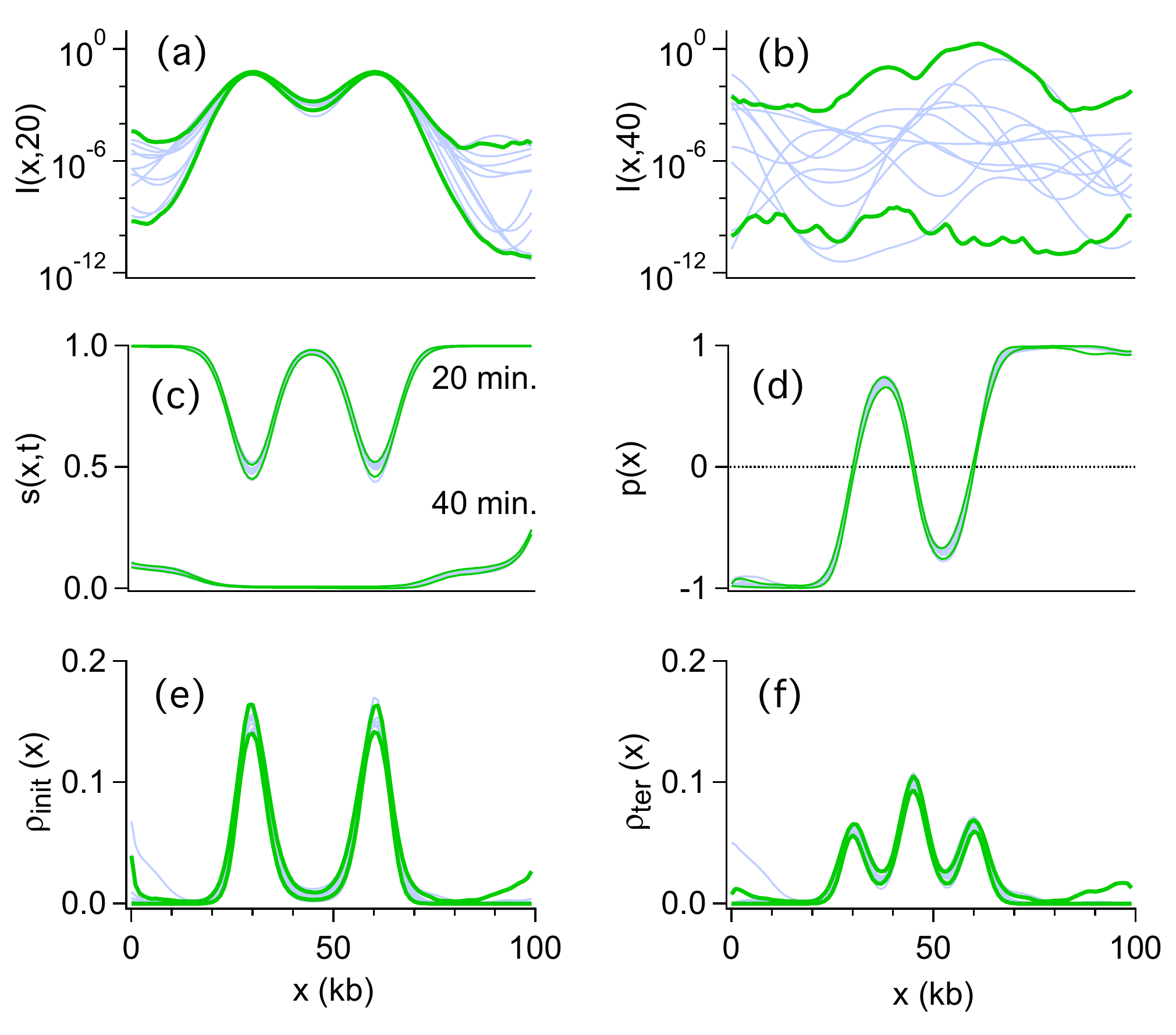}
  \caption{
    \label{fig:inference}
    (Color online) 
    Negligible posterior uncertainty, except for the initiation rate at the end of S-phase.
    Ten MCMC samples (light blue lines) from the posterior probability distribution 
    Eq.~(\ref{eq:Bayes}), 
    and the $90\%$ credible interval (between heavy green lines).
    (a) Local initiation rate at $t=20$ min. and 40 min (b).   (c)
    Unreplicated fractions at $t=20$ and 40 min. (d) Replication fork polarity.  
    (e) Spatial density of initiation and termination (f).
  }
\end{figure}

To estimate the width of the posterior distribution, Eq.~\eqref{eq:Bayes}, we used Markov Chain Monte Carlo sampling.  We first implemented the classic  Metropolis-Hastings algorithm, but it was very slow.  We then tried instead the Hamiltonian Monte Carlo algorithm \cite{MacKay2003}, which was about 100 times faster.  We initialized the Markov chain at the MAP estimate in order to skip the burn-in phase and used the Hessian of the energy $E(m)$ as a preconditioning matrix for the momentum.  We generated an effectively independent sample (i.e., an evaluation over the entire spacetime grid) every 5 seconds on a regular laptop.  Ten samples from the posterior distribution are given in Fig.~\ref{fig:inference}, as well as the $90\%$ credible interval.  We see that the posterior uncertainty for the unreplicated fraction, Fig.~\ref{fig:inference}(c), the replication fork polarity, Fig.~\ref{fig:inference}(d), the density of initiation, Fig.~\ref{fig:inference}(e), and termination, Fig.~\ref{fig:inference}(f),  are negligible, with a small posterior uncertainty at the boundaries.  The local initiation rate has low posterior uncertainty, Fig.~\ref{fig:inference}(a), except at the end of S-phase, Fig.~\ref{fig:inference}(b).  The large uncertainty on the initiation rate at the end of S-phase is easily understandable:  At the end of S-phase, the unreplicated fractions are close to zero; thus, even large variations of the local initiation rate result in minor variations in the unreplicated fractions that will be much smaller than the noise level.  The initiation rate thus cannot be accurately inferred in these regions.  However, as we have seen in Fig.~\ref{fig:inference}(c)-(f), the large uncertainty in the local initiation rate at the end of S-phase results in negligible uncertainty for other quantities of interest.

\subsubsection{Inferring $v$  and $\ell_0$}

We inferred the most important hyperparameters, the fork velocity $v$ and the spatial smoothness scale $\ell_0$, directly from the data.  By Bayes' theorem applied to the hyperparameters in Eq.~\eqref{eq:hyperparam}, the posterior distribution for $v$ and $\ell_0$ is given by
\begin{equation}
	P( v, \ell_0 | d, \beta') 
		= \frac{1}{P(d | \beta')} \, P( d | v, \ell_0, \beta')  \, P( v, \ell_0 ) \,,
\end{equation}
where $\beta' = \{ \sigma_d, I_0, \sigma_0, \tau_0 \}$ contains the remaining hyperparameters.  If we assume a flat prior on $v$ and $\ell_0$, the posterior $P( v, \ell_0 | d, \beta')$ is simply proportional to the evidence $P( d | v, \ell_0, \beta') = P(d | \beta)$.  From Eq.~\eqref{eq:evidence}, the evidence $P(d | \beta)$ is evaluated by integrating the joint posterior $P(m,d | \beta)$ over $m$, a $10^4$-dimensional vector.  Such a high-dimensional integration cannot be performed numerically.  In the Laplace approximation \cite{MacKay2003}, the joint posterior is approximated by a Gaussian around its maximum (the MAP estimate $m_{\rm MAP}$):
\begin{equation}
	P( m , d | \beta) \simeq e^{- E_{\rm MAP} 
	- \frac{1}{2} (m-m_{\rm MAP}) \cdot E_{\rm MAP}^{''} (m-m_{\rm MAP})} \,,
\label{eq:Laplace_approx}
\end{equation}
where $E_{\rm MAP}$ is the energy Eq.~(\ref{eq:energy}) at the MAP, and $E_{\rm MAP}^{''}$ is the Hessian of the energy evaluated at the MAP.  As the distribution is a Gaussian, the integration over $m$ can be done analytically.   The log evidence is then 
\begin{equation}
	\ln P(d | \beta) \simeq \tfrac{1}{2} \ln \det(2 \pi E_{\rm MAP}^{''}) - E_{\rm MAP} \,.
\label{eq:Laplace_approx_evidence}
\end{equation}
This formula corresponds to the saddle-point approximation often encountered in statistical physics.

We then evaluated the Laplace approximation of the log evidence on a grid of values for $v$ and $\ell_0$, spanning $v=1$ kb/min to $v=3$ kb/min every 0.1 kb/min for the fork velocity, and $\ell_0=0$ kb to $l=20$ kb every 5 kb for the spatial smoothing scale.  We found that the value of the evidence at $v=2$ kb/min and $\ell_0 = $ 15 kb (the true values of the artificial data set) was several orders of magnitude larger than  the evidence at other values.  In other words, the posterior probability for $v$ and $\ell_0$ is, at the resolution considered, almost equal to one at the true values of  $(v,\ell_0)$ and zero elsewhere.  Therefore, for the data set considered here, we can infer accurately (at a resolution of 0.1 kb/min and 5 kb) the fork velocity and the spatial scale with near certainty.

\section{Conclusions}

In this article, we have generalized the forward analysis of the DNA replication problem to the case of arbitrary initiation rates $I(x,t)$.  We then introduced an inference procedure based on a Gaussian process prior that avoids the need of earlier curve-fitting methods to specify the form of $I(x,t)$ in advance.  We then showed that a small test case (100 kb genome) with typical replication parameters and typical experimental noise and resolution could be successfully inverted, with very small errors for all replication quantities of interest, except in cases where the experimental data were not very informative.  (These cases were typically the end of S phase and the edges of the sample.) The method may in principle be generalized to handle realistic genome sizes. 

Assuming that the method does scale up and can successfully reproduce earlier analyses, we will then have a powerful method for learning about DNA replication in multiple organisms.  Further, while we have focused on microarray and sequencing experiments, our methods should be compatible with the numerous other experimental methods, including fluorescence-activated cell sorting (FACS)\cite{ma12}, molecular combing \cite{ma12}, and Okazaki-fragment mapping \cite{mcguffee13}.  Moreover, while the analysis is conceptually more complicated than curve fitting, it can be automated and thus has the potential to be more widely used in the biological community.

From a more theoretical point of view, Gaussian process regression \cite{Rasmussen2006} can be regarded as the equivalent of a free field theory, in that the objects of interest are fields (defined over space and time) and are supposed to always show Gaussian fluctuations.  In our case, the nonlinear relation between the replication data and the initiation rate of interest meant that our result was far from Gaussian.  Although we used MCMC methods to sample the resulting non-Gaussian distributions, it would be interesting to explore other approaches to data analysis.  In one approach, the parameter space of the probabilistic model defines a Riemannian  manifold, allowing one to formulate a search algorithm for the MAP estimate \cite{transtrum10,*transtrum11} or MCMC exploration \cite{girolami11} in geometric terms.  Taking a geometric approach can speed up the numerical algorithms discussed here.  Alternatively, one can use the equivalent of interacting field theories and not assume Gaussian distributions.  In this regard, the work of En{\ss}lin and collaborators on \textit{information field theory} \cite{oppermann13} is an especially promising approach.


\begin{acknowledgments}

We thank Scott Yang and Nick Rhind for their suggestions.  This work was funded by NSERC (Canada).
\end{acknowledgments}

\appendix*

\section{}
\label{sec:proof}

We prove Eqs.~\eqref{eq:ds_dt}-\eqref{eq:d2T_dx2} by first considering the replication program in one cell cycle. Then we show that the results derived for a single cell cycle generalize straightforwardly to the ensemble average for a stochastic or variable replication program. 

\subsection{In one cell cycle}
\label{sec:in1cycle}

Consider $N$ origins $O_1, \ldots, O_N$ located at genomic positions $x_1 < \ldots < x_N$ and initiated at times $t_1,\ldots , t_N$, with fork velocities $\pm v$.  From simple geometry [Fig.~\ref{fig:replication_program}(a)], we see that each pair of origins $(O_i, O_{i+1})$ leads to a single termination event $\Lambda_i$ at location $x_i^{(\Lambda)}$ and time $t_i^{(\Lambda)}$, where
\begin{align}
	x_i^{(\Lambda)} &= \tfrac{1}{2} (x_{i+1}+x_i) + \tfrac{1}{2}v(t_{i+1}-t_i) \,, 
      		\nonumber \\
	t_i^{(\Lambda)} 
      		&= \tfrac{1}{2v}(x_{i+1} - x_i) + \tfrac{1}{2}(t_{i+1}+t_i) \,.
\end{align}
The spatiotemporal densities of initiation and termination are therefore
given by
\begin{align}
	\label{eq:rho_ini}
	\rho_{\rm init}(x,t) &= \sum_i \delta(x-x_i)\delta(t-t_i) \,, \nonumber \\  
	\rho_{\rm ter}(x,t) &= \sum_i \delta \left( x-x_{i}^{(\Lambda)} \right) 
		\delta \left(t-t_{i}^{(\Lambda)} \right) \,,
\end{align}
where $\delta (\cdot)$ is the Dirac delta function.  Integrating over time gives the corresponding spatial densities:
\begin{equation}
	\rho_{\rm init}(x) = \sum_i \delta(x-x_i),  \quad 
	\rho_{\rm ter}(x)= \sum_i \delta \left(x-x_{i}^{(\Lambda)} \right) \,.
\end{equation}

The replication timing curve $T(x)$ is defined as the time at which the locus
$x$ is replicated and is represented as the solid line in  Fig.~\ref{fig:replication_program}(a).  Let us define the \textit{domain of origin $O_i$} to be $x \in \bigl[x_{i-1}^{(\Lambda)} ,x_{i}^{(\Lambda)} \bigr]$.  Within the domain, the replication timing curve is given by
\begin{equation}
	T(x) = t_i + \frac{| x - x_i |}{v} \,.
\label{eq:T-near-ori}
\end{equation}
The straight lines about each origin are one-dimensional analogs of the ``light cones'' of relativity that radiate from a source.  In the similarly defined \textit{domain of terminus~$\Lambda_i$}, defined as $x \in [x_{i-1},x_i]$ and illustrated in Fig.~\ref{fig:replication_program}(a), the replication timing curve is given by the  ``past cones'' from $\Lambda_i$:
\begin{equation}
	T(x) = t_{i}^{(\Lambda)} - \frac{ \left| x - x_{i}^{(\Lambda)} \right|}{v} \,.
\label{eq:T-near-ter}
\end{equation}
The \textit{unreplicated fraction} $s(x,t)$ is given by 
\begin{equation}
	s(x,t) = H[T(x)-t] \,,
\end{equation}
where $H$ is the Heaviside step function.
 
In Fig.~\ref{fig:replication_program}(a) in the domain of origin $O_i$, right-moving and left-moving replication forks have densities that are given by
\begin{equation}
  	\rho_{\pm}(x,t) = H[\pm(x-x_i)] \, \tfrac{1}{v} \delta [t - T(x)] \,,
\end{equation}
Equivalently, in the domain of terminus $\Lambda_i$, the fork densities are given by
\begin{equation}
  	\rho_{\pm}(x,t) = H[\mp(x-x_{i}^{(\Lambda)})] \, \tfrac{1}{v} \delta [t - T(x)] \,,
\end{equation}
Note that $p_\pm (x) = \int_0^{t_\infty} v dt \, \rho_{\pm} (x,t)$ equals 1 if the locus $x$ is replicated by a $\pm$ fork.  Thus, the replication fork polarity $p(x) = p_+(x) - p_-(x) = \pm 1$ gives the directionality $(\pm)$ of the fork replicating the locus $x$.  In the domain of origin $O_i$, the replication fork polarity is equal to
\begin{equation}
	p(x) = \mathrm{sign}(x-x_i) \,.
\label{eq:p}
\end{equation}

It is then straightforward, using the theory of distribution \cite{Rudin1991} and the above definitions Eqs.~\eqref{eq:rho_ini}-\eqref{eq:p}, to differentiate $s(x,t)$ and check the relations Eqs.~\eqref{eq:ds_dt}-\eqref{eq:d2T_dx2}. 

\vspace{1em}

\subsection{Ensemble average}
\label{sec:ensemble-avg}

Because of the stochasticity of the replication program \cite{aFriedmanGenesCells1997,aPatelMolBiolCell2006,aRhindNatCellBiol2006,aCzajkowskyJMolBiol2008}, the number of activated origins, their positions, and their firing times, all change from one cell cycle to another [Fig.~\ref{fig:replication_program}(b)].  This variability may also reflect heterogeneity in the population of cells considered.  For instance, mixtures of different cell types or cells of the same cell type but with different epigenetic states can give different stochastic replication programs.  The ensemble average then corresponds to a superimposition of the different replication programs.  A clear-cut example of the latter is the replication program in the human female $X$ chromosome, where the ensemble average of replication seems to be ``biphasic,'' superposing the replication programs from the  active and inactive $X$ chromosomes \cite{aHansenProcNatlAcadSciUSA2010,koren14}.

The unreplicated fraction $s(x,t)$, the densities of initiation $\rho_{\rm init}(x,t)$ and termination $\rho_{\rm ter}(x,t)$, the fork densities $\rho_{\pm}(x,t)$, the fork polarity $p(x)$ and the mean replication timing $T(x)$ defined in Sec.~\ref{sec:definitions} all correspond to the ensemble averages of their one-cell-cycle counterparts given in Sec.~\ref{sec:in1cycle}. We proved in Sec.~\ref{sec:in1cycle} that the relations Eqs.~\eqref{eq:ds_dt}-\eqref{eq:d2T_dx2} were true in each cell cycle. As derivatives and averages commute, we can straightforwardly extend Eqs.~\eqref{eq:ds_dt}-\eqref{eq:d2T_dx2} to the ensemble average.

\bibliographystyle{BIB/apsrev4-1}
\bibliography{inference_kjma}


\end{document}